\documentclass[a4paper,twocolumn,11pt,unpublished]{quantumarticle}
\pdfoutput=1
\usepackage[utf8]{inputenc}
\usepackage[english]{babel}
\usepackage[T1]{fontenc}
\usepackage{amsmath}
\usepackage{hyperref}

\usepackage{tikz}
\usepackage{lipsum}

\usepackage{cite}
\usepackage{amsmath,amssymb,amsfonts}
\usepackage{algorithmic}
\usepackage{algorithm}

\usepackage{subcaption} % or subfig, depending on your preference
\usepackage{graphicx}
\usepackage{textcomp}
\usepackage{xcolor}
\usepackage{braket}
\usepackage{tabularx}
\usepackage{qcircuit}  
\usepackage{orcidlink}
\usepackage{booktabs}

\begin{document}

\title{Multi-Target Rydberg Gates via Spatial Blockade Engineering}
\author{Samuel Stein\orcidlink{0000-0002-2655-8251}}
\affiliation{Future Computing Research Group, Pacific Northwest National Laboratory, USA}
\email{samuel.stein@pnnl.gov}

\author{Chenxu Liu\orcidlink{0000-0003-2616-3126}}
\affiliation{Future Computing Research Group, Pacific Northwest National Laboratory, USA}

\author{Shuwen Kan\orcidlink{0009-0004-0726-5260}}
\affiliation{Future Computing Research Group, Pacific Northwest National Laboratory, USA}
\affiliation{Department of Computer and Information Science, Fordham University}

\author{Eleanor Crane\orcidlink{0000-0002-2752-6462
}}
\affiliation{Department of Physics, King’s College London, Strand, London, WC2R 2LS, UK}
\author{Yufei Ding\orcidlink{0000-0002-8716-5793
}}
\affiliation{Department of Computer Science \& Engineering, University of California San Diego}

\author{Ying Mao\orcidlink{0000-0002-4484-4892}}
\affiliation{Department of Computer \& Information Science, Fordham University}

\author{Alexander Schuckert\orcidlink{0000-0002-9969-7391}}
\affiliation{Joint Quantum Institute and Joint Center for Quantum Information and Computer Science, University of Maryland and NIST, USA}
\email{alexander@schuckert.org}

\author{Ang Li\orcidlink{0000-0003-3734-9137}}
\affiliation{Future Computing Research Group, Pacific Northwest National Laboratory, USA}
\affiliation{Department of Electrical \& Computer
Engineering, University of Washington, USA}

\maketitle
\begin{abstract}
Multi-target gates offer the potential to reduce gate depth in syndrome extraction for quantum error correction. Although neutral-atom quantum computers have demonstrated native multi-qubit gates, existing approaches that avoid additional control or multiple atomic species have been limited to single-target gates. We propose single-control-multi-target $\mathrm{C}(Z^{\otimes N})$ gates on a single-species neutral-atom platform that require no extra control and have gate durations comparable to standard $\mathrm{C}Z$ gates. Our approach leverages tailored interatomic distances to create an asymmetric blockade between the control and target atoms. Using a GPU-accelerated pulse synthesis protocol, we design smooth control pulses for $\mathrm{C}ZZ$ and $\mathrm{C}ZZZ$ gates, achieving fidelities of up to $99.55\%$ and $99.24\%$, respectively, even in the presence of simulated atom placement errors and Rydberg-state decay. This work presents a practical path to implementing multi-target gates in neutral-atom systems, significantly reducing the resource overhead for syndrome extraction.

\end{abstract}

\section{Introduction}

Quantum computing research has advanced to the point where many error-correction experiments have been demonstrated in several platforms~\cite{google2023suppressing,ai2024quantum,reichardt2024demonstration,livingston2022experimental,reichardt2024logical,bluvstein2024logical,rodriguez2024experimental,henriet2020quantum,xu2024constant,tan2022qubit,graham2022multi,evered2023high,baranes2025leveraging}. However, achieving large-scale fault-tolerant operation remains challenging, with a key obstacle in neutral-atom platforms being the long clock time determined by the measurement of multi-qubit observables of the form $\hat Z^{\otimes N}$, where $N=4$ for the surface code and $N=6$ for the triangular color code. Typically, these measurements are performed by applying a roughly depth-$N$ circuit which maps this observable to the measurement of $\hat Z$ on an ancilla. An alternative approach would be the application of a $N$-qubit single-control-multi-target gate of the form ($\mathrm{C}(Z^{\otimes N})$)~\cite{devitt2013quantum}, which could not only reduce the clock time of the logical processor but also reduce errors such as atom heating induced by moving atoms in-between gates. In neutral atoms, the Rydberg blockade~\cite{Lukin2001} enables multi-qubit gates, but most naturally, a permutationally symmetric $(\mathrm{C}^{\otimes N})Z$ gate is realized~\cite{jandura2022time,evered2023high}. Proposals for multi-target gates, which singles out the control atom from the targets, so far involve additional experimental controls such as local microwave control or additional lasers compared to the conventional single-species, single-Rydberg-state blockade mechanism used for standard $\mathrm{C}Z$ gates~\cite{wu2010implementation,su2018rydberg,su2017fast,muller2009mesoscopic,young2021asymmetric}. %Among the competing hardware approaches, neutral atom architectures show particular promise, offering large qubit numbers and low control overheads enabling straight-forward transversal gate implementation~\cite{h}.

\begin{figure}
    \centering
    \includegraphics[width=1\linewidth]{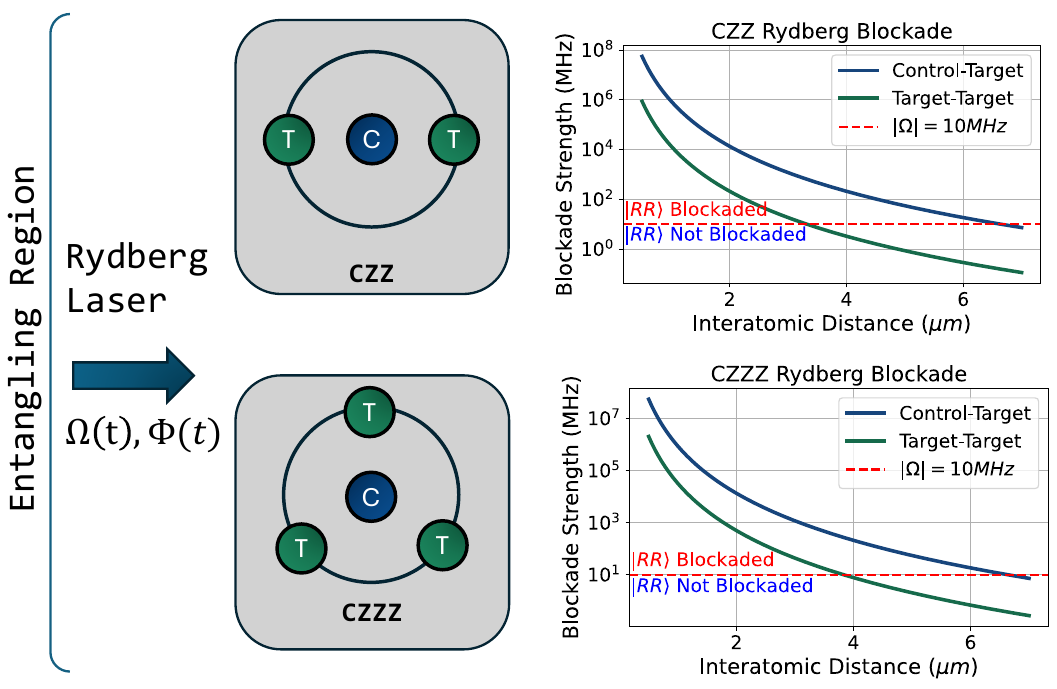}
    \caption{Geometric arrangements and asymmetric blockade engineering for multi-qubit gates. (Top Row) A symmetric $\mathrm{CC}Z$ gate relies on placing all atoms close together for mutual Rydberg blockade \cite{evered2023high}. For the asymmetric $\mathrm{C}ZZ$ (N=2 targets) and $\mathrm{C}ZZZ$ (N=3 targets) gates proposed here, the single control atom (C) is placed at the center, while the N target atoms (T) are distributed evenly on the perimeter of a 2D circle around the control. (Bottom Row Plots) Calculated blockade strength versus interatomic distance for Control-Target (C-T) and Target-Target (T-T) pairs in the $\mathrm{C}ZZ$ and $\mathrm{C}ZZZ$ configurations. This geometry ensures shorter C-T distances (strong blockade) compared to T-T distances (weak blockade). Blockade strengths below the drive Rabi frequency |$\Omega$| (dashed line, e.g., 10MHz) are considered negligible. Asymmetric blockade, required for $\mathrm{C}(Z^{\otimes N})$ gates, emerges when the T-T is below $|\Omega|$ while the C-T strength remains well above it.}
    \label{fig:begin_fig}
\end{figure}

In this work, we show that an asymmetric blockade for realizing $\mathrm{C}ZZ$ and $\mathrm{C}ZZZ$ gates can be achieved through purely geometric positioning on single-species neutral atom hardware, hence removing the requirements of additional laser or microwave fields. We propose an arrangement where one control atom strongly blockades multiple target atoms, while target–target blockade is kept negligible, by placing the control at the center and targets on a perimeter, c.f. Fig.~\ref{fig:begin_fig}, and then numerically engineering an intricate pulse sequence. This approach aims to retain the operational simplicity of using a single global drive pulse, similar to standard $\mathrm{C}Z$ operations \cite{evered2023high}.% The experimental complexity is thus shifted from intricate laser control schemes or specialized interaction regimes towards the precise and stable spatial arrangement of the atoms.

Specifically, we numerically determine the pulse sequence required to implement $\mathrm{C}ZZ$ and $\mathrm{C}ZZZ$ gates with high-fidelity on timescales comparable to today’s two-qubit gates. We examine the robustness of this geometric approach, accounting for realistic noise channels including Rydberg state decay and atom position fluctuations. In a companion work~\cite{stein2025}, we show that $\mathrm{C}ZZ$ and $\mathrm{C}ZZZ$ multi-qubit gates achieve the optimal depth for readout of the surface code and color code stabilizers. Hence, we expect our work to enable a depth-optimal readout of stabilizers in topological codes in neutral atoms. %Achieving these performance targets enables multi-qubit $\mathrm{C}(Z^{\otimes N})$ gates $(N=2,3)$ in neutral atom platforms without substantial additional control complexity, thus potentially reducing algorithmic clock time, atom reconfiguration demands, and streamlining operations such as syndrome extraction~\cite{nielsen2010quantum}.

\section{Background}
\label{sec:background}
\subsection{Neutral Atom Quantum Computing}

Neutral atom systems employ laser cooling and optical trapping techniques to confine individual atoms in highly ordered arrays. The qubit is typically encoded in long-lived hyperfine or electronic ground state, with single-qubit gates implemented via direct laser pulses on individual atoms. Multi-qubit gates are realized by coupling qubits to a highly excited Rydberg state using a global laser drive on the $\ket{1} \leftrightarrow \ket{r}$ transition \cite{evered2023high}. By temporarily exciting one or more qubits into a Rydberg state, strong atom-atom interactions emerge, creating a “blockade” region that is energetically blockaded from multiple simultaneous Rydberg excitations within a certain volume. This effect underpins multi-qubit entangling operations between atoms that lie within the blockade radius \cite{wilk2010entanglement,saffman2010quantum,saffman2016quantum}. 

One of the appealing features of neutral atom quantum computing is the ability to rearrange and scale qubit arrays with relative ease by shifting optical traps. Large qubit registers with flexible geometries have been demonstrated \cite{wurtz2023aquila}. Single-qubit gate fidelities are extremely high, with reported error rates on the order of $10^{-5}\%$-$10^{-3}\%$, whereas current two-qubit gate implementations show error rates around $0.5\%$. Measurement operations have also been demonstrated with typical error rates on the order of $0.5\%$~\cite{evered2023high}.

Neutral atom platforms have demonstrated remarkable promise,  with clear pathways to scaling out to larger systems \cite{henriet2020quantum}, road maps towards fault tolerance \cite{auger2017blueprint}, and experiments even demonstrating logical computation \cite{reichardt2024logical} and magic state distillation \cite{rodriguez2024experimental}.

\subsection{Rydberg Blockade Mechanism}
\label{sec:Rydberg_blockade}

The Rydberg blockade is central to neutral-atom quantum computing, enabling multi-qubit entanglement in a single step via strong dipole-dipole interactions. When an atom is excited to a high-lying Rydberg state $\ket{r}$, nearby atoms experience a large energy shift if they also attempt to occupy $\ket{r}$. Provided the interaction $|V_{ij}|$ far exceeds the Rabi frequency $\Omega$, only one atom in the blockade region can be excited at a time~\cite{saffman2010quantum,urban2009observation}. 
The blockade strength typically decays as $d^{-6}$ with inter-atomic separation $d$, making atom placement and laser parameters crucial for maintaining high-fidelity operations.

Considering an array of atoms, driven by an on-resonant laser to excite the atoms from the $|1\rangle$ state to the $|r\rangle$ state, the atom Hamiltonian in the rotating-wave frame can be written as \cite{jandura2022time}:

\begin{equation}
\label{eq:H_multi_atom_merged}
\begin{aligned}
    H(t) \;=\;& \sum_{j=1}^n
        \frac{\Omega_{j}(t)}{2} \ket{r}_j\bra{1}_j + h.c\\
      &+ \sum_{j<k} V_{jk}\,\ket{r}_j \ket{r}_k \bra{r}_j \bra{r}_k,
\end{aligned}
\end{equation}
where $V_{jk}$ is the dipole-dipole interaction energy between atoms $j$ and $k$ when both are in $\ket{r}$~\cite{jandura2022time}. The atoms are $|1\rangle_j$ and $|r\rangle_j$ are coupled by the laser with time-dependent Rabi frequency $\Omega(t)=|\Omega_j(t)| e^{i\phi( t)}$. $\Omega(t)$ is taken to be complex, with $|\Omega _j(t)|$ encoding amplitude and $e^{i\phi (t)}$ encoding phase. We absorbed the detuning of the Rydberg laser into the time-dependence of the phase.  $V_{i,j}$ represents the inter-atomic blockade strength, and is highly dependent on inter-atomic distances. If $|V_{ij}|\gg \Omega$ for all pairs within the blockade radius, population of doubly-excited Rydberg states is energetically suppressed, ensuring that at most one $\ket{r}$ excitation occurs (“blockaded”). Atoms located outside the blockade radius experience negligible mutual interaction and can be excited simultaneously if desired \cite{jandura2022time}. 

In the simplest two-qubit scenario, this blockade-induced nonlinearity directly yields a \textit{controlled-Z} ($\mathrm{C}Z$) gate. With atoms $i$ and $j$ in the strong-blockade regime ($|V_{ij}|\gg \Omega$), the doubly excited state $\ket{r}_i\ket{r}_j$ is 
energetically inaccessible. Second-order perturbation theory, treating  \(\ket{r}_i\ket{r}_j\) as off resonant, shows that in the subspace $\{\ket{10}, \ket{01}, \ket{11}\}$, an effective interaction $\tfrac{\Omega^2}{4\,V_{ij}}\,\ket{11}\bra{11}$ appears. Driving the 
$\ket{1}\!\leftrightarrow\!\ket{r}$ transition for a time $t$ satisfying 
\(
    \tfrac{\Omega^2}{4V_{ij}}\,t \;=\; \pi
\)
applies a $\pi$ phase only to the $\ket{11}$ component, implementing a $\mathrm{C}Z$ gate 
(up to single-qubit phases)~\cite{evered2023high,jandura2022time}. In many systems, the pulse is applied globally to all atoms within an entangling region, offering a path toward scalable, parallel entangling operations \cite{evered2023high}. Achieving high gate fidelity requires fine-tuned pulse shaping and precise atom placement to maintain $|V_{ij}|\gg \Omega$ where needed, while also minimizing spurious interactions between distant qubits.

\subsection{Atom Rearrangement}

Neutral-atom quantum processors offer a unique capability: they can physically rearrange qubits (i.e. individual atoms) during a computation to achieve flexible connectivity. However, this reconfigurability comes at a significant cost. Rearrangement significantly increases both the overall computation time and the likelihood of atom loss. Moving a single atom via optical tweezers typically requires between hundreds to tens of thousands of microseconds  \cite{wang2024atomique,ebadi2021quantum}, while a two-qubit entangling $\mathrm{C}Z$ gate can be executed on the order of a few hundred nanoseconds \cite{evered2023high}. This disparity spans several orders of magnitude in time.

Frequent shuttling of atoms not only adds significant latency to the overall circuit run-time but also introduces additional error channels. Atom loss and decoherence can worsen significantly, given the finite lifetime of optical traps. Although neutral atoms have exceedingly long lifespans when compared to gates,even these long lifetimes can be compromised by the durations required for reconfiguration \cite{ebadi2021quantum}.

Because rearrangement is so time-consuming and error-prone \cite{ebadi2021quantum}, high-depth routines such as repeated syndrome extraction become prohibitively expensive. Thus, reducing rearrangement overhead whilst preserving high-fidelity gates remains crucial.

\subsection{Impact of $\mathrm{C}(Z^{\otimes N})$ gates on Error Correction}

A primary motivation for developing multi-target gates is their potential to streamline quantum error correction (QEC) protocols. A key step in QEC is syndrome extraction, which involves measuring multi-qubit stabilizer operators to detect errors~\cite{devitt2013quantum, nielsen2010quantum}. Many important QEC codes, such as surface codes~\cite{fowler2012surface} or color codes~\cite{Bombin2006Topological}, utilize stabilizers involving interactions between multiple data qubits (e.g., weight-4 or higher). Color codes are especially prominent in the neutral atom field, with breakthroughs in neutral atoms utilizing the color code \cite{rodriguez2024experimental,reichardt2024logical}.

Standard approaches typically decompose the measurement of a weight-$k$ stabilizer into a sequence of $k$ individual two-qubit entangling gates (like $\mathrm{C}Z$) involving an ancilla qubit~\cite{nielsen2010quantum}. This inherently requires at least $k$ entangling layers in the circuit. Executing these layers sequentially can contribute significantly to the circuit depth and may necessitate substantial atom rearrangement in architectures like neutral atoms, increasing execution time and error susceptibility~\cite{xu2024constant, wang2024atomique}.

Multi-target gates, such as the C($Z^{\otimes N}$) gates developed here, offer an alternative paradigm. For a stabilizer measurement involving $k$ data qubits, a C($Z^{\otimes k-1}$) or similar multi-target gate could potentially implement the required collective phase logic in fewer entangling layers. This reduction in entangling layers could significantly decrease circuit depth and potentially alleviate demands on atom rearrangement.

\section{Methods: Pulse Optimisation of Asymmetric Rydberg gates}
\label{sec:noise_robust_pulses}

We target a single global pulse on single-species hardware similar to that in Ref. \cite{wurtz2023aquila}, implementing $\mathrm{C}(Z^{\otimes N})$ gates using similar settings to those used for global $\mathrm{C}Z$ gates.

\subsection{\texorpdfstring{\(\mathrm{C}(Z^{\otimes N})\)}{C(Z^N)} Gates Under Asymmetric Rydberg Blockade}
\label{sec:proof_cZtensorN}

Many previous neutral-atom gate designs focus on \(\mathrm{C}^{\otimes N}Z\) (multi-control \(Z\)) operations~\cite{evered2023high}, whereby a $-1$ phase is applied to homogeneous $|11..1\rangle$ states. While such gates are often useful, many multi-qubit operations instead require a single control qubit that applies phase shifts to multiple targets. An illustrative example is the two-target case, \(\mathrm{C(Z^{\otimes 2})}\), which assigns distinct phases to basis states like \(\ket{101}\) compared to \(\ket{011}\). By contrast, a \(\mathrm{CCZ}\) gate (i.e.\ two-control, one-target) acts non-trivially only on \(\ket{111}\) and thus cannot implement these different phase evolutions.
However, realizing \(\mathrm{C}(Z^{\otimes N})\) gates on neutral atoms is non-trivial due to the nature of the Rydberg blockade. If all atoms lie within a single blockade radius and one single laser drive is used, any multi-\(\ket{1}\) configuration couples \emph{symmetrically} to the Rydberg state, preventing distinct phases from being applied to different patterns of excitation. Consequently, a simple all-in-one-blockade arrangement cannot yield \(\mathrm{C}(Z^{\otimes N})\). 
We propose to break this symmetric coupling by spatially arranging qubits such that the control-target and target-target interactions differ. In particular, we place the control qubit at the center of a circle, with each target positioned sufficiently far apart on the perimeter so that:

\begin{itemize}
    \item \textbf{Control--target blockade:} The distance \(d_{0j}\) between the control and each target is small enough to ensure a large van der Waals interaction \(V_{0j} \gg \Omega\), establishing a strong blockade for each control--target pair.
    \item \textbf{Weak target--target coupling:} Any two targets \(j, k\) are separated sufficiently so that \(V_{jk}\) is negligible, allowing them to be simultaneously excited without undesired cross-blockade.
\end{itemize}

This arrangement suppresses target-target blockade while ensuring the control-target blockade remains strong, thus enabling distinct phase accumulations required by $\mathrm{C}(Z^{\otimes N})$.

\subsection{Atom Placement for \texorpdfstring{$\mathrm{C}(Z^{\otimes N})$}{C(Z^N)} Gates}

A key step in implementing multi-target gates such as \(\mathrm{C}(Z^{\otimes N})\) is choosing a geometric layout of the atoms that ensures strong blockade between the control qubit and each target, but sufficiently negligible blockade among the targets themselves. One effective approach is to place the control atom at the center of a circle and distribute the \(N\) target atoms uniformly around the perimeter. Thus,
\begin{itemize}
    \item The control atom is at the origin, providing shortest distances (and therefore strong dipole-dipole interactions) to all targets.
    \item The targets are as far apart from one another as possible, reducing inter-target interactions to an insignificant level.
\end{itemize}

The control atom is located at the origin, and we place each target atom \(j\) (\(j=1,\dots,N\)) at
\[
\mathbf{x}_j
\;=\;
\bigl(R\cos\theta_j,\,
      R\sin\theta_j,\,
      0\bigr),
\quad
\theta_j
\;=\;
\frac{2\pi}{N}\,j,
\]
for some chosen radius \(R\). The \emph{control--target} blockade effect is then
\[
V_{0j}
\;=\;
\frac{C_6}{\|\mathbf{x}_0-\mathbf{x}_j\|^6}
\;=\;
\frac{C_6}{R^6},
\]
where \(\mathbf{x}_0=(0,0,0)\) denotes the control atom’s location. For an appropriately chosen \(R\), we have \(V_{0j}\gg \Omega\), ensuring strong blockade between the control qubit and any target, and \(V_{jk}\approx 0\), ensuring the targets do not blockade each other.  We still consider the 3D arrangement, because noise randomly perturbs along the $z$ axis aswell.

Formally, any two targets \(j\neq k\) on the circle are separated by
\[
d_{jk}
\;=\;
\bigl\|\mathbf{x}_j-\mathbf{x}_k\bigr\|
\;=\;
2\,R\,
\sin\!\Bigl(\tfrac{\pi}{N}\Bigr).
\]
To keep the interaction \(V_{jk} = C_6 / d_{jk}^6\) negligible compared to the Rabi frequency \(\Omega\), one must choose \(R\) large enough so that
\[
\frac{C_6}{\bigl(2\,R\,
\sin(\tfrac{\pi}{N})\bigr)^6}
\;\;\ll\;\;\Omega.
\]
Hence, pushing the targets radially outward can break the target--target symmetry whilst preserving sufficiently strong control--target blockade while keeping target--target interactions low. However, as \(N\) increases, maintaining these conditions grows more 
challenging, as more target atoms are placed on the perimeter. For instance, a $\mathrm{C}(Z^{\otimes 4})$ gate exhibits the limit of a reduced target-target blockade of $\frac{C_6}{8R^6}$ against the target-control blockade $\frac{C_6}{R^6}$, making the system increasingly sensitive to atom motion and demanding more complex pulse shaping. At small \(N\), such as a $\mathrm{C}(Z^{\otimes 2})$(two targets), choosing \(R\) is much less sensitive, as the relative target-target blockade is $\frac{C_6}{64R^6}$. As \(N\) grows and \(R\) must be increased, \(\lvert V_{0j}\rvert = 
C_6 / R^6\) diminishes, and the system breaks down and the arrangement no longer satisfies the required blockade hierarchy once target-target distances are less than that of control-target. 
In simulation, we find that beyond $\mathrm{C}(Z^{\otimes 2})$ or $\mathrm{C}(Z^{\otimes 3})$, the added complexity and diminishing control-target blockade strength for larger N in this specific circular geometry make achieving high fidelity under realistic noise conditions increasingly difficult with a single global pulse. This geometric constraint inherently limits the direct applicability of this simple circular arrangement to N=2 and N=3 target gates in practice for current experimental parameters and desired fidelities. With lower variance in atom positioning, $\mathrm{C}(Z^{\otimes 4})$ and $\mathrm{C}(Z^{\otimes 5})$ may be possible, as they maintain target-target distances less than control-target. Higher weight operators are not possible in our proposed setup due to target-target distances being less than or equal to the target-control distance.

\subsection{Noise Channels and Atomic Perturbations in Neutral Atom Hardware}
\label{sec:noise-and-perturbations}

High-fidelity multi-qubit gates on neutral-atom hardware must account for a variety of imperfections, including spontaneous decay from the Rydberg state, realistic control dynamics, and small perturbations of atoms from their ideal positions in 3D space. In our approach, we incorporate all of these noise sources into a numerical pulse-optimization routine that aims to generate pulses robust against hardware-induced deviations and are well-suited to near-term control hardware.

We model two dominant categories of imperfections: (i) \emph{Rydberg decay} via local amplitude-damping–type Kraus maps, and (ii) small random displacements in the atomic positions (in all three spatial dimensions) that modify the inter-atomic interaction strength. 

\paragraph{Decay/Amplitude Damping}
With entangling gates being on the order of hundreds of nanoseconds, and logical states $|0\rangle,|1\rangle$ having second-long lifespans, thermal decay of these states is near zero. However, the Rydberg state has a finite lifespan and is a dominant source of error in entangling gates \cite{evered2023high}. 

To model this noise channel, at each short time step \(\Delta t\), after evolving under the system Hamiltonian, we apply a local Kraus channel that transfers population from the excited Rydberg state \(\ket{r}\) to the ground states \(\ket{0}\) and \(\ket{1}\), and leakage out of the computational subspace to $|e\rangle$. We do not track amplitudes of $|e\rangle$, and model this as amplitude dampening. Concretely, for each qubit \(i\), we define 
\[
    p \;=\; \gamma \,\Delta t,
\]
where \(\gamma\) is the decay rate, and \(p\) is partitioned into three decay probabilities \(p_0\), \(p_1\),  and \(p_2\) for the channels \(\ket{r}\rightarrow\ket{0}\), \(\ket{r}\rightarrow\ket{1}\) and \(\ket{r}\rightarrow\ket{e}\)  respectively (with \(p_0+p_1+p_2 =p\)). The associated Kraus operators acting on the three-level system \(\{\ket{0},\,\ket{1},\,\ket{r}\}\) are given by
\begin{align}
E_0 \;=\;
\begin{pmatrix}
1 & 0 & 0 \\[1mm]
0 & 1 & 0 \\[1mm]
0 & 0 & \sqrt{1-p}
\end{pmatrix}, 
\quad
E_1 \;=\;
\begin{pmatrix}
0 & 0 & \sqrt{p_0} \\[1mm]
0 & 0 & 0 \\[1mm]
0 & 0 & 0
\end{pmatrix} \\[2mm]
E_2 \;=\;
\begin{pmatrix}
0 & 0 & 0 \\[1mm]
0 & 0 & \sqrt{p_1} \\[1mm]
0 & 0 & 0
\end{pmatrix}
\end{align}
Modeling the \(\ket{r}\rightarrow\ket{e}\) transition comprises omitting the $p_2$ kraus channel, and represents the amplitude lost from $|r\rangle$ to $|e\rangle$. This amplitude is permanently lost. In this way, population decays from the Rydberg state \(\ket{r}\) into both the computational basis states as well as a leakage state, thus approximating a continuous amplitude-damping process.  

\paragraph{Atomic Position Perturbations.}

The Rydberg blockade is highly sensitive to interatomic spacing, and small positional errors along all three spatial directions can substantially degrade gate fidelity. Atoms exhibit systematic trap errors, resulting in deviations in atom positioning (i.e., atom-placement errors) that arise from imperfections in the trapping potential. In addition to these systematic errors, inherent atomic motion arises both from photon recoil during Rydberg excitation and decay and from the intrinsic thermal and quantum zero‐point fluctuations present even in an ideal harmonic trap \cite{wurtz2023aquila,pagano2022error}.

To account for all these contributions in our optimization procedure, at each optimizer iteration we draw random displacements 
\[
\boldsymbol{\delta}_i = (\delta x_i,\delta y_i,\delta z_i)
\]
for each atom \(i\). These displacements update the nominal coordinates as
\[
\mathbf{r}_i \rightarrow \mathbf{r}_i + \boldsymbol{\delta}_i.
\]
thereby modifying the pairwise blockade interaction
\[
    V_{ij} \;=\; \frac{C_6}{\|\mathbf{x}_i + \boldsymbol{\delta}_i - (\mathbf{x}_j + \boldsymbol{\delta}_j)\|^6}.
\]
The displacements \(\delta x_i\), \(\delta y_i\), and \(\delta z_i\) are sampled from a gaussian distribution with widths $\sigma_x,\sigma_y$, and $\sigma_z$, allowing the optimizer to learn a more robust gate solution that tolerates typical three-dimensional positional variations. We optimize to maximizing the average fidelity across these samples:
\[
\overline{\mathcal{L}} \;=\; \frac{1}{N_{\mathrm{shots}}}\,\sum_{m=1}^{N_{\mathrm{shots}}} 
  \bigl[1 - \mathcal{F}_{m}\bigr],
\]
where \(\mathcal{F}_{m}\) is the fidelity for the \(m\)-th displacement sample. This "mini-batch” approach yields control pulses that maintain high fidelity over typical atom-motion distributions. 

\subsection{Gate Pulse Synthesis}

\subsubsection{Base Pulse Synthesis}
We target the generation of time-dependent Rabi pulses \(\Omega(t)\) with a phase profile $\Phi(t)$  that maximizes a gate fidelity metric:
\[
\mathcal{F}\Bigl(U_{\mathrm{target}},\,\mathcal{E}_{\mathrm{Real}}(\Omega,\Phi)\Bigr).
\]
Where $U_{target}$ is our target gate, and $\mathcal{E}_{Real}$ is the learned evolution over input basis states $\{{|0\rangle,|1\rangle,|+\rangle,|i\rangle}\}^{\otimes N}$.
We discretize the total evolution time \(T\) into \(n\) small segments, each of duration \(\Delta t = T/n\). In segment \(k\), the Hamiltonian is parameterized by \(\Omega_k\) and \(\Phi_k\). We then perform a trotterized evolution for each segment:
\[
    U_k \;=\; \exp\bigl(-\mathrm{i}\,\Delta t\,H_k(\Omega_k,\Phi_k)\bigr),
\]
where 
\[
    H_k \;=\; \frac{\Omega_k}{2}\bigl(e^{\mathrm{i}\phi_k}D_{+} + e^{-\mathrm{i}\phi_k}D_{-}\bigr)\; \;+\; \sum_{i<j}V_{ij}\,n_i n_j,
\]
and \(D_{\pm}\) are the collective raising/lowering operators \(\ket{1}\!\leftrightarrow\!\ket{r}\), while \(n_i\) projects onto the Rydberg state of atom \(i\). After each segment, we apply the Kraus operators \(\{E_0,E_1,E_2\}\) to model amplitude damping. For a given pulse schedule \(\{\Omega_k,\phi_k\}\), the final operation is
\[
U_{\mathrm{real}} \;=\; \Bigl(\prod_{k=1}^{n} K(\Delta t) U_k\Bigr)
\]
with \(K(\Delta t)\) denoting the combined effect of Kraus operations in each sub-step.

Due to its success in related quantum-control tasks \cite{song2022optimizing,leng2023efficient}, we use the Adam optimizer \cite{kingma2014adam}. The base loss function is set to maximize fidelity, i.e. $1-\mathcal{F}(U_{target},\mathcal{E}_{Real})$.

\subsubsection{Ensuring Realistic Pulse Shapes}

High-fidelity solutions found via unconstrained optimization can exhibit 
rapid, large-amplitude variations in their pulse profiles, making them 
difficult or impossible to implement on real hardware. To address this, 
we include a smoothness penalty that discourages abrupt changes in 
amplitude or phase between consecutive time segments. Specifically, we 
define
\begin{equation}
    \gamma_i = \bigl|\Phi_{i+1} - \Phi_i\bigr|,\quad i = 1,\ldots,n-1,
    \label{eq:phase_diff}
\end{equation}
where \(\Phi_k\) represents either the phase or amplitude at time slice~\(k\). 
We then introduce the cost function

\begin{equation}
\begin{split}
\text{C}_{\mathrm{smooth}} &= \sum_{i=1}^{n-1} f(\gamma_i ), \quad \text{where} \\
f(d_i) &= \begin{cases}
       \lambda_s\, \gamma_i ^2, & \text{if } \gamma_i  \le \tau, \\[1mm]
       \lambda_s\, \tau^2 + \lambda_b\Bigl(e^{B\,(\gamma_i -\tau)} - 1\Bigr), & \text{if } \gamma_i  > \tau.
    \end{cases}
\end{split}
\label{eq:threshold_penalty_def}
\end{equation}
which penalizes small parameter jumps quadratically, and large jumps exponentially. 
During the early stages of training, we keep $\lambda_s$ high initially, and reduce to 0 over the first half of training. We still penalize large discontinuities via \(\lambda_{b}\). 

\begin{figure}[htbp]
    \centering
    \begin{subfigure}[b]{0.23\textwidth}
        \centering
        \includegraphics[width=\textwidth]{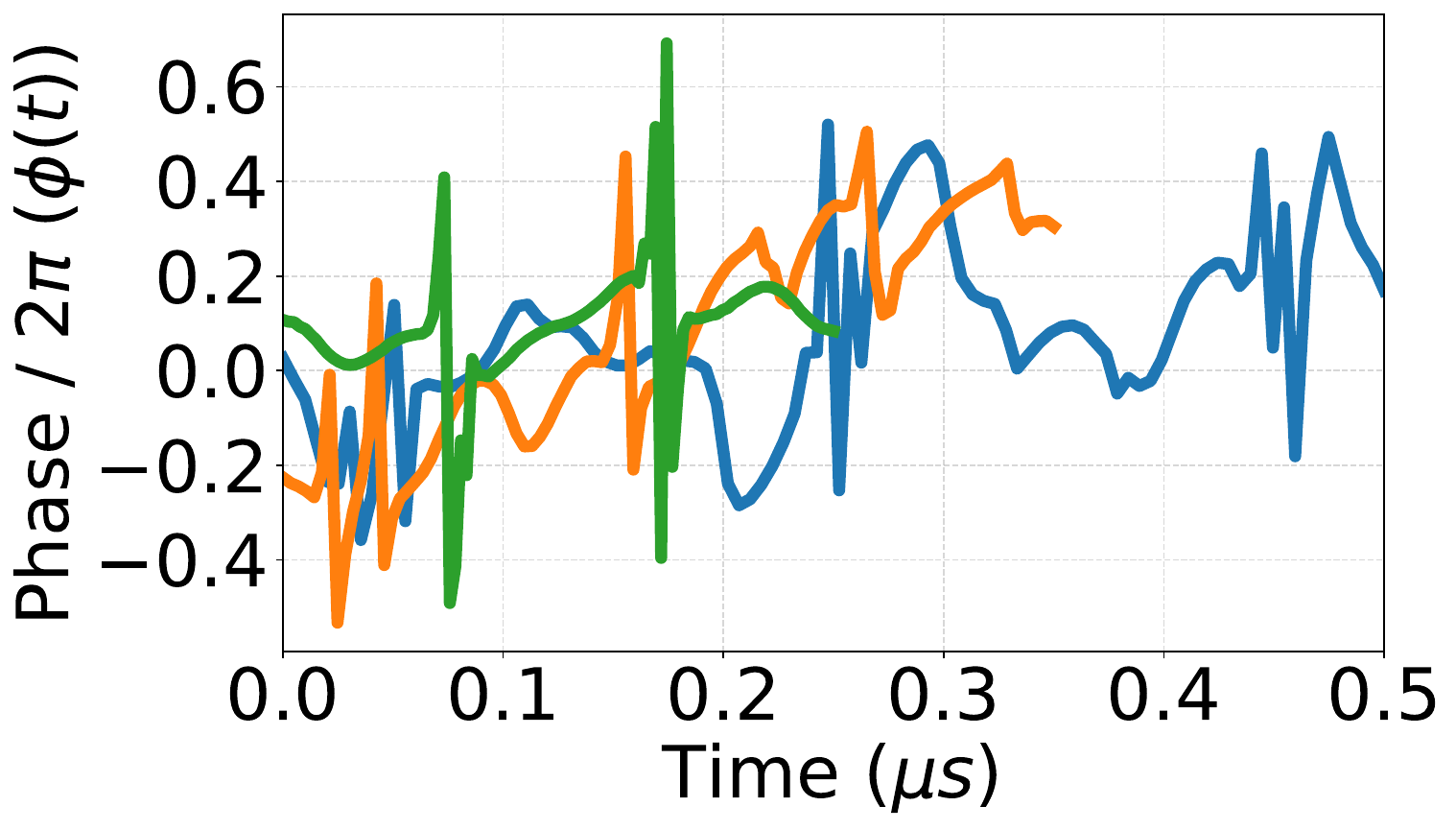}
        \caption{Without equation \ref{eq:threshold_penalty_def}}
        \label{fig:no_smoothing}
    \end{subfigure}
    \hfill
    \begin{subfigure}[b]{0.23\textwidth}
        \centering
        \includegraphics[width=\textwidth]{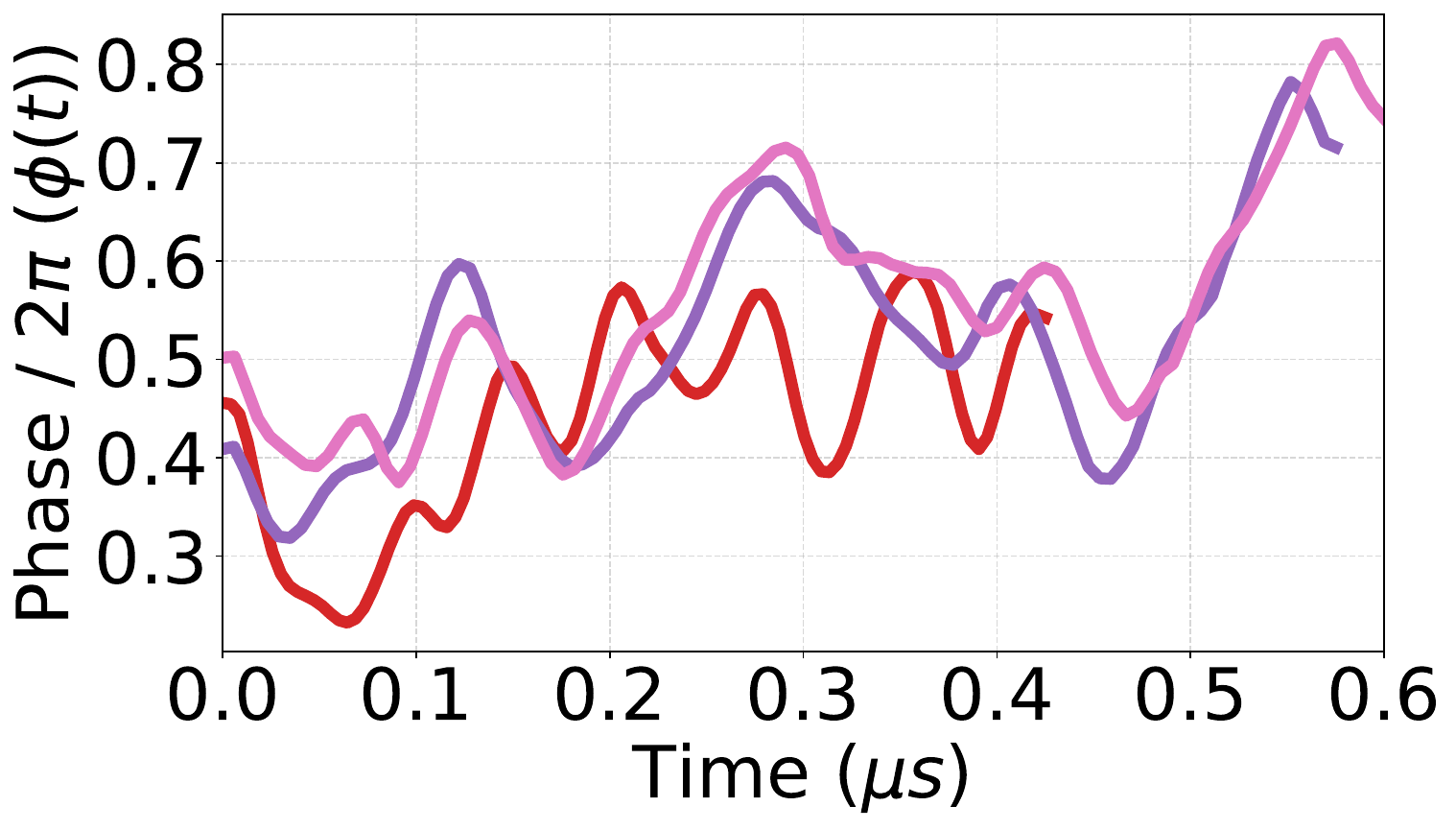}
        \caption{With equation \ref{eq:threshold_penalty_def}}
        \label{fig:smoothing}
    \end{subfigure}
    \caption{Two sets of 3 samples of pulses from training. Each line is an individual high fidelity pulse for a $\mathrm{C}(Z^{\otimes 2})$ gate, and have no relation to each other. (a) has no smoothing penalty and (b) with the smoothing penalty included.  (a) shows examples without the smoothing term. In contrast, (b) shows the loss function enforces smooth pulses by penalizing large phase jumps, thereby discouraging unrealistic trajectories}
    \label{fig:training_examples}
\end{figure}

Figure~\ref{fig:training_examples} illustrates two sets of converged pulses with 
and without this smoothing term. The penalty effectively encourages the optimizer to yield solutions that are both high fidelity and smooth, thereby being implementable on real hardware. By balancing hardware feasibility with pulse-design freedom, our smoothness cost ensures that the final schedules achieve strong performance 
without relying on nonphysical control trajectories.

Parameters are set based on experimental stability, and refined in practice. In actual experiments, these values might be further fine-tuned to match hardware constraints.

\section{Results: $\mathrm{C}(Z^{\otimes N})$ Gates}
\subsection{Pulse Generation for \texorpdfstring{$\mathrm{C}(Z^{\otimes N})$}{C(Z^N)}: Speed Limit Under Ideal Conditions}
\label{sec:pulse_generation_ideal}

We begin by benchmarking and validating our proposed gate sets are numerically viable, and the potential atom coordinates that enable them. We characterize the \(\mathrm{C}(Z^{\otimes N})\) gate set in an ideal scenario, with no dynamic atom motion, control constraints, or thermal decay. We establish a baseline for these gates being possible, and how quickly these gates can be performed while still achieving high fidelity. 

We begin by setting a fixed Rabi frequency, \(\Omega_{\max}= 10\times 2\pi\,\mathrm{MHz}\). Simultaneously, we explore various control-target radii, varying the Rydberg blockade strength both between control-target, and target-target. This is captured in the radius parameter. Ideally, control-target is $\gg\Omega$ and target-target is $\ll\Omega $, though this is not physically realistic and trade offs exist in the radii of the system.

We provide a given total gate duration \(T\), and discretize the evolution into \(n=100\) segments, each characterized by a constant drive parameter \(\{|\Omega_{Max}|\}\) and the learnable parameter \(\{\Phi\}\). 
\[
\mathcal{F}\Bigl(\mathcal{E}_{\mathrm{Real}},\,\mathrm{C}(Z^{\otimes N})\Bigr).
\]
Each optimization run evaluates the trotterized evolution and backpropagates the fidelity gradient over the control parameters. By sweeping across different fixed values of \(T\), over $50$ random initializations to avoid converging in a local minima, and various radii, we can identify atom placements that are capable of high fidelity $\mathrm{C}(Z^{\otimes N})$ gate and their minimum durations.

\begin{figure}[t]
    \centering
    % Replace the filename below with the actual name of your figure file.
    \includegraphics[width=1\linewidth]{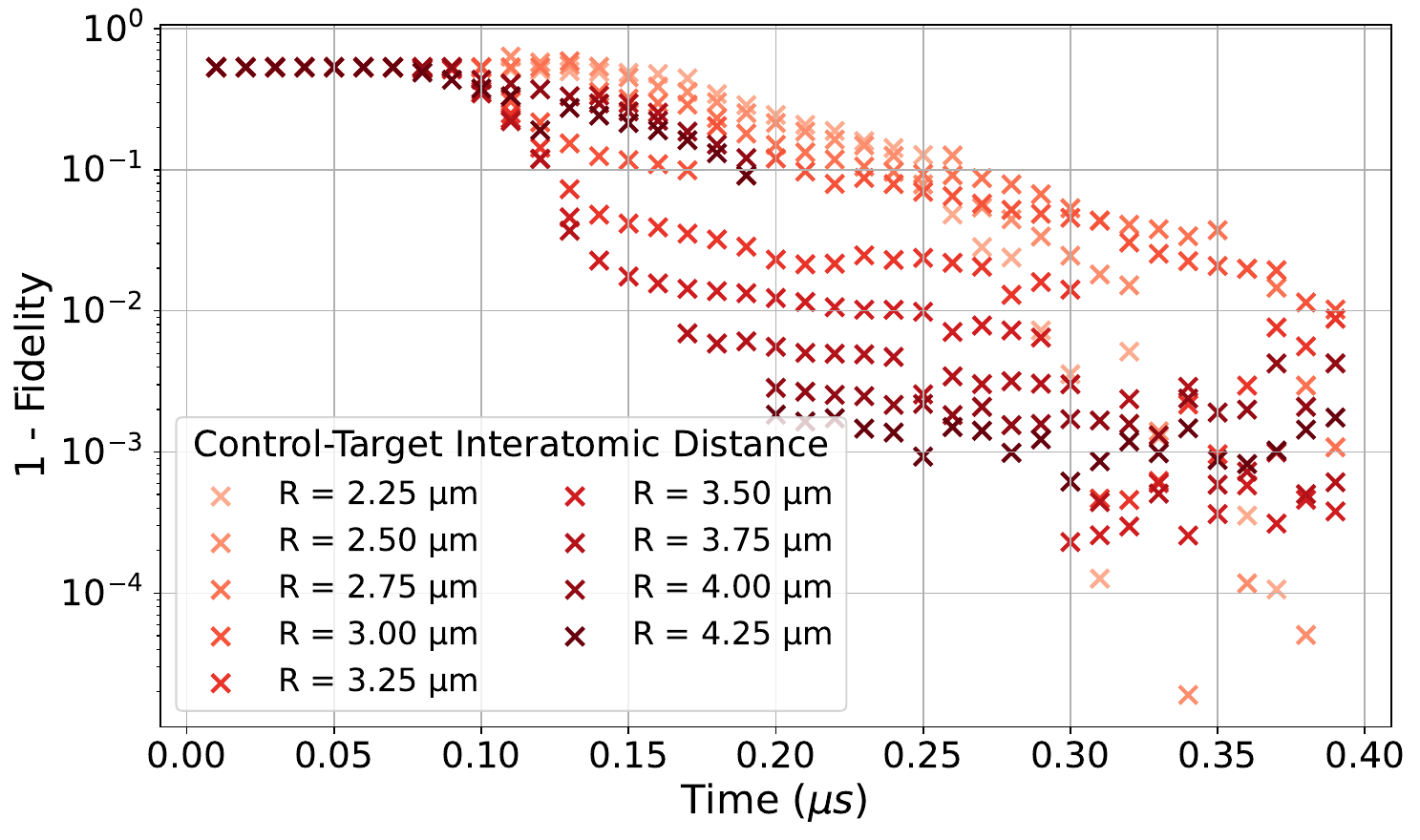}
    \caption{
        Achieved fidelities for a $\mathrm{C}(Z^{\otimes 2})$ gate under ideal conditions (no atomic motion, no decay), plotted against the total evolution time \(T\). Each color/marker denotes a distinct atom-to-control radius \(R\), ranging from $2.25\mu m$ to $4.25\mu m$. For intermediate radii ($3.25\mu m, 3.50\mu m,3.75\mu m$), the gate rapidly attains high fidelity within $0.15\mu s$. Smaller or larger radii require longer  durations to surpass 80\% fidelity.}
    \label{fig:CZZ_ideal}
\end{figure}

\begin{figure}[t]
    \centering
    % Replace the filename below with the actual name of your figure file.
    \includegraphics[width=1\linewidth]{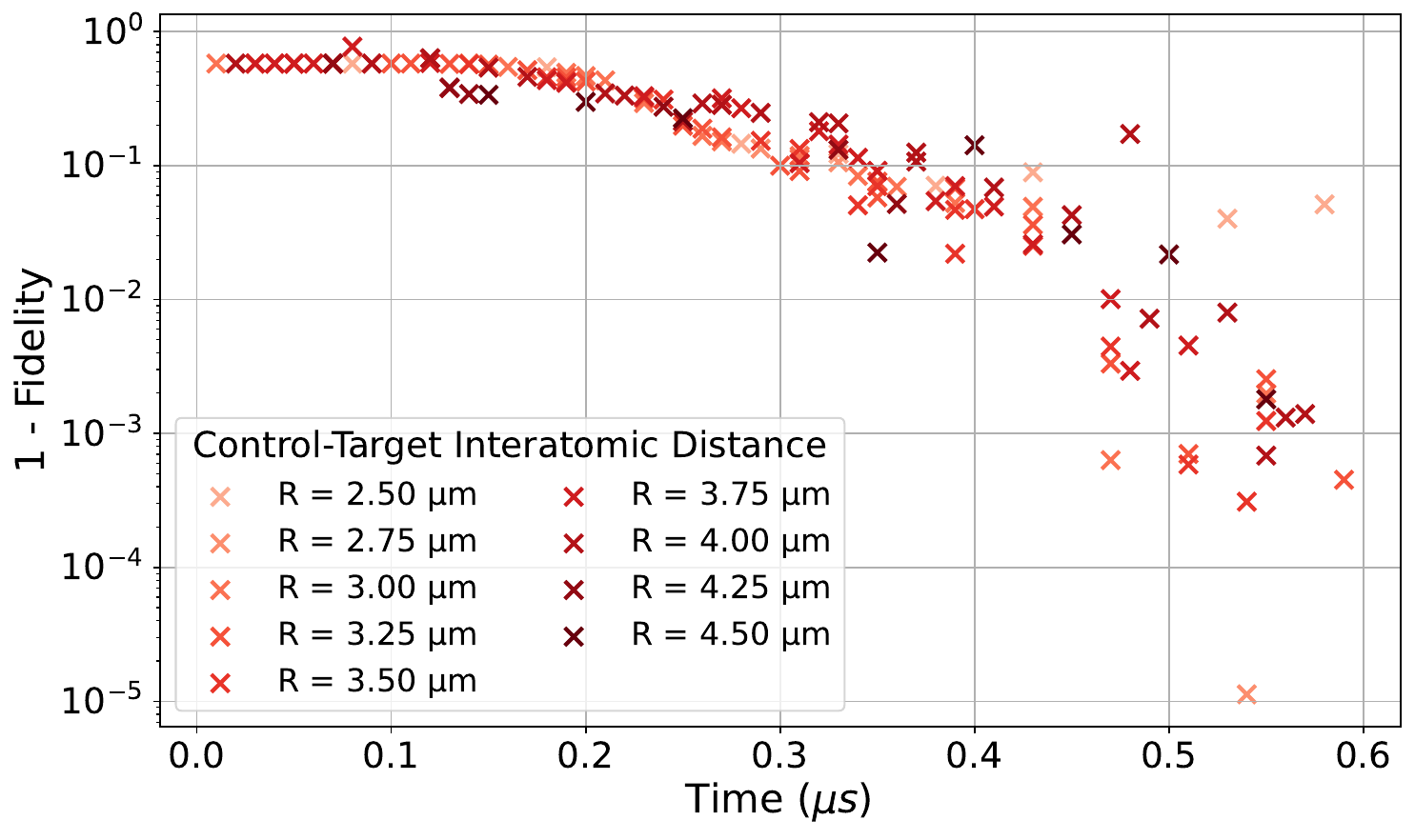}
    \caption{
        Achieved fidelities for a $\mathrm{C}(Z^{\otimes 3})$ gate under ideal conditions (no atomic motion, no decay), plotted against the total evolution time \(T\). Each color/marker denotes a distinct atom-to-control radius \(R\), ranging from $2.5\mu m$ to $4.5\mu m$. No clear good interatomic distance emerges, and slower gate times when compared to Figure \ref{fig:CZZ_ideal} are required to implement the $\mathrm{C}(Z^{\otimes 3})$ gate. 
        }
    \label{fig:CZZZ_ideal}
\end{figure}

In \figurename~\ref{fig:CZZ_ideal}, we illustrate the final fidelity of a $\mathrm{C}(Z^{\otimes 2})$ gate 
for various inter-atomic distances \(R\). Each data set was obtained by optimizing based on the parameters discussed prior under Equation \ref{eq:H_multi_atom_merged}. The rapid fidelity increase seen 
for certain intermediate radii between $3.25\mu m$ and $3.75\mu m$ indicates a performant atom configuration where the control-target blockade is sufficiently large, yet target-target blockade remains sufficiently suppressed. In contrast, excessively small or large radii force the optimizer to either extend the gate duration 
or accept lower fidelity outcomes. This analysis confirms that even in a noiseless scenario, choosing an appropriate distance \(R\) impacts our gates performance significantly, and more importantly, the $\mathrm{C}(Z^{\otimes 2})$ gate is possible at high fidelity. 

Meanwhile, in Figure \ref{fig:CZZZ_ideal} we show our simulations for a $\mathrm{C}(Z^{\otimes 3})$ gate under various evolution times and radii. Here, no clear “optimal” radius emerges, and longer gate times are required to attain high fidelity solutions. This underscores the added geometric and timing constraints of a three-target blockade arrangement, wherein cross-talk between control atoms grows more significant and saturates the achievable fidelity for short pulses. However, we find that high fidelity solutions at fidelities $>99.9\%$ do exist for $\mathrm{C}(Z^{\otimes 3})$ gates, with Radii between $3-4 \mu m$ demonstrating the best fidelities found, aligning well with current neutral atom hardware atom positioning parameters \cite{evered2023high,wurtz2023aquila}.

\begin{table}[h!]
\centering
\caption{Final Parameters Used in Pulse Training and Experiments}
\label{tab:params}
\renewcommand{\arraystretch}{1.2}
\begin{tabularx}{0.5\textwidth}{l X}
\hline
\textbf{Parameter} & \textbf{Value} \\
\hline
\textbf{Rydberg Lifetime} 
  & 88\,\(\mu\mathrm{s}\)  \cite{evered2023high} \\[5pt]

\textbf{Atomic Species}
  & Rb \\[5pt]

\textbf{Small Jump Penalty} 
  & \(\lambda_s = 0.01 \to 0\)  \\[5pt]

\textbf{Large Jump Penalty} 
  & \(\lambda_b = 1.0\) \\[5pt]

\textbf{Exponent Factor} 
  & \(B = 2.0\) \\[5pt]

\textbf{Gate Duration Range} 
  & \(T = 250 \sim 600\,\mathrm{ns}\) \\[5pt]

\textbf{Global Rabi Drive} 
  & \(\Omega_{\max}=2\pi\times10\,\mathrm{MHz}\) \\[5pt]

\textbf{$\mathrm{C}_6$ Rydberg Blockade} 
  & \(862690 \times 2\pi \text{ } \mathrm{MHz} \text{ }\mu \mathrm{m}^6 \) \cite{wurtz2023aquila}\\[5pt]

\textbf{Ramp‐Up Time} 
  & 10\,ns \\[5pt]

\textbf{Displacement Samples} 
  & \(10{,}000\) \\[5pt]

\textbf{Kraus P Values} 
  & \(p_0 = 0.1354,\, p_1= 0.2504,\, p_2=0.6142\) \cite{evered2023high} \\[5pt]
  
\textbf{Atom Motion (x,y)} 
  & \(\sigma_x=\sigma_y=0.014\mu \mathrm{m}\) \cite{pagano2022error,chew2022ultrafast} \\[5pt]

\textbf{Atom Motion (z)} 
  & \(\sigma_z=0.16\,\mu\mathrm{m}\) \cite{pagano2022error,chew2022ultrafast} \\
\hline
\end{tabularx}
\end{table}

\begin{table}[ht]
    \centering
    \caption{Fidelity, gate time, and control-target radius of various learned pulses. 
             The reported fidelity is the average over random displacements 
             described in Table~\ref{tab:params}.}
    \label{tab:refined_fidelity}
    
    \renewcommand{\arraystretch}{1.2}  % Increase vertical spacing
    \setlength{\tabcolsep}{12pt}       % Increase horizontal spacing
    
    \begin{tabularx}{0.5\textwidth}{l *{3}{>{\centering\arraybackslash}X}}
        \toprule
        \textbf{Gate} & \textbf{Fidelity} & \textbf{Time (ns)} & \textbf{Radius ($\mu$m)} \\
        \midrule
        $\mathrm{C}(Z^{\otimes 2})$  & 0.9955 & 575 & 3.5 \\
        $\mathrm{C}(Z^{\otimes 2})$  & 0.9913 & 400 & 3.5 \\
        $\mathrm{C}(Z^{\otimes 2})$  & 0.9880 & 375 & 3.5 \\
        $\mathrm{C}(Z^{\otimes 2})$  & 0.9821 & 325 & 3.5 \\
        $\mathrm{C}(Z^{\otimes 3})$ & 0.9924 & 575 & 4.0 \\
        $\mathrm{C}(Z^{\otimes 3})$ & 0.9821 & 525 & 4.0 \\
        $\mathrm{C}(Z^{\otimes 3})$ & 0.9818 & 575 & 3.5 \\
        \bottomrule
    \end{tabularx}
\end{table}

\subsection{Understanding Noise Channels}

To understand where the dominant sources of noise persist in our ideal pulses, we make use of a Pauli Transfer Map (PTM) \cite{greenbaum2015introduction,hantzko2024pauli}. PTMs map pauli operators to pauli operators, allowing us to identify unwanted channels within a gate. We compare our learned unitaries to the ideal comparable $\mathrm{C}(Z^{\otimes N})$ gate. 

We analyze our pulses with $>99\%$ fidelity from Figure \ref{fig:CZZ_ideal} and \ref{fig:CZZZ_ideal}. We observe that the key error channels are:
\begin{itemize}
    \item ${X}\leftrightarrow Y$ coupling predominantly on the control qubit, with lower weight coupling on the target qubits, resulting in coherent phase errors on the control.
    \item $Z\leftrightarrow I$ coupling, from residual Rydberg population.
\end{itemize}
At lower radii ($2.0\mu \mathrm{m}$), the dominant noise channels are a coherent $X\leftrightarrow Y$ error on the control qubit, indicating a phase error. Another dominant error source is represented by an error channel between ${Z\leftrightarrow I}$. These couplings represent amplitude dampening in the form of residual Rydberg population. When we evaluate on Pauli Transfer Maps, we only consider the states $|0\rangle$ and $|1\rangle$, hence residual population in the $|r\rangle$ state is equivalent to amplitude decay.
At higher radii ($>4.0\mu \mathrm{m}$) we observe a similar pattern, however with $I\leftrightarrow Z$ errors being the dominant source of error. We observe a consistent error structure in the Pauli Transfer Map, with residual $|r\rangle$ population and coherent errors being the dominant persistent errors.

We visualize the average Pauli Transfer Map for the ideal and learned $\mathrm{C}(Z^{\otimes 2})$ gate under Appendix \ref{app:A}. 

The PTM highlights the two primary physical processes limiting fidelity in this scheme: coherent phase accumulation errors likely arising from imperfect pulse shaping (manifesting as X$\leftrightarrow$Y), and incoherent population loss from residual Rydberg state population (manifesting as Z$\leftrightarrow$I). Understanding the relative contribution of these distinct physical error mechanisms is imperative for tailoring pulse sequences to mitigate the dominant error source for specific parameter regimes.

\begin{figure*}[t]
    \centering
    %
    % -- Duplicate for surface code scheduling figure --
    \begin{subfigure}[b]{0.32\textwidth}
        \centering
        \includegraphics[width=\textwidth]{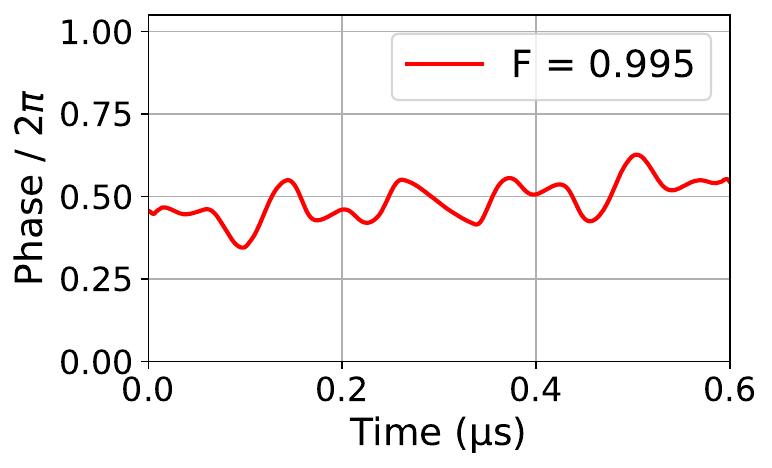}
        \caption{}
        \label{fig:pulse1_czz}
    \end{subfigure}
    \hfill
    \begin{subfigure}[b]{0.32\textwidth}
        \centering
        \includegraphics[width=\textwidth]{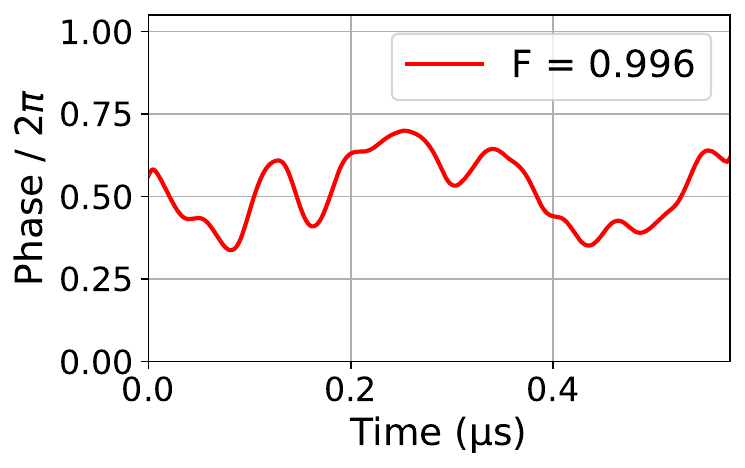}
        \caption{}
        \label{fig:pulse2_czz}
    \end{subfigure}
    \hfill
    \begin{subfigure}[b]{0.32\textwidth}
        \centering
        \includegraphics[width=\textwidth]{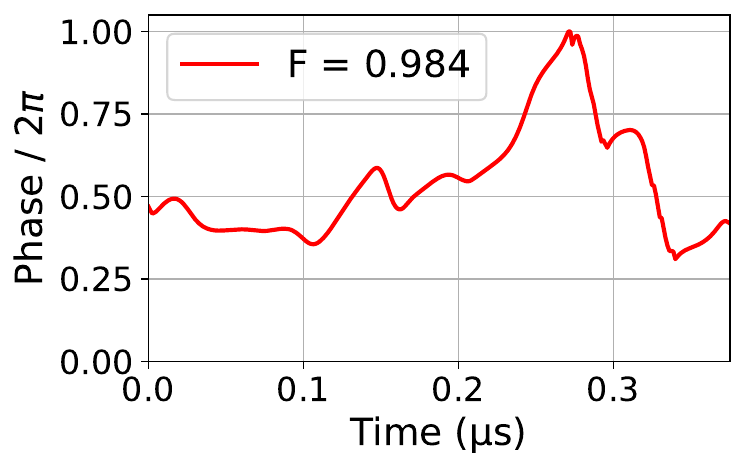}
        \caption{}
        \label{fig:pulse3_czz}
    \end{subfigure}
    \caption{
      Pulses for $\mathrm{C}(Z^{\otimes 2})$ with Fidelity reported. Each pulse is trained and evaluated with thermal noise and atom motion}
    \label{fig:czz_results_real}
\end{figure*}

\begin{figure*}[t]
    \centering
    %
    % -- Duplicate for surface code scheduling figure --
    \begin{subfigure}[b]{0.32\textwidth}
        \centering
        \includegraphics[width=\textwidth]{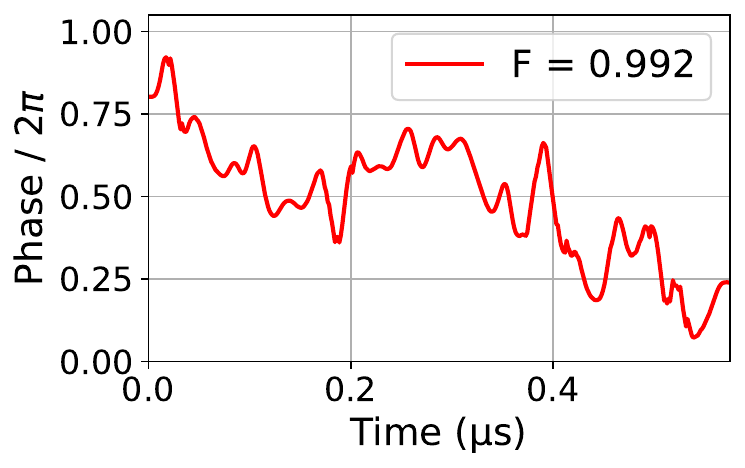}
        \caption{}
        \label{fig:pulse1_czz}
    \end{subfigure}
    \hfill
    \begin{subfigure}[b]{0.32\textwidth}
        \centering
        \includegraphics[width=\textwidth]{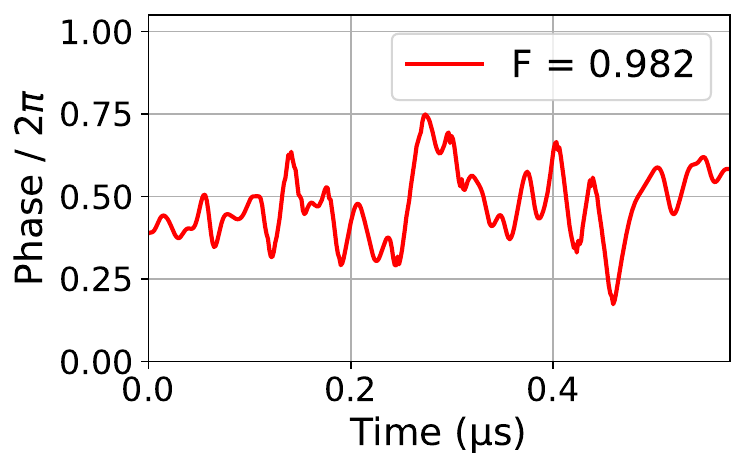}
        \caption{}
        \label{fig:pulse2_czz}
    \end{subfigure}
    \hfill
    \begin{subfigure}[b]{0.32\textwidth}
        \centering
        \includegraphics[width=\textwidth]{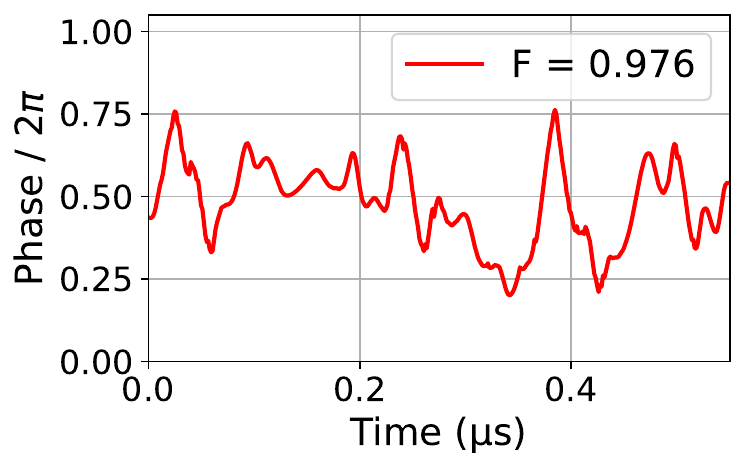}
        \caption{}
        \label{fig:pulse3_czz}
    \end{subfigure}
    \caption{
      Pulses for $\mathrm{C}(Z^{\otimes 3})$ with fidelity reported. Each pulse is trained and evaluated with noise and atom motion}
    \label{fig:czzz_results_real}
\end{figure*}

\subsection{Realistic Pulse Generation with Noise and Atom Motion}
\label{sec:realistic_pulse_generation}

While the idealized pulse designs of Section~\ref{sec:pulse_generation_ideal} provide a lower bound on gate durations and potential fidelities, real hardware faces additional noise mechanisms and control constraints that can significantly degrade performance. In particular, neutral-atom devices exhibit atom-position fluctuations, spontaneous decay out of Rydberg states, and pulse shaping constraints. We now extend our pulse-optimization approach to account for these challenges, using a robust training procedure that simultaneously samples thousands of perturbed Hamiltonians and penalizes large, rapid changes in pulse parameters. This procedure ensures that our multi-target Rydberg gates remain realistic and high fidelity under realistic conditions.

We randomly sample 3D displacements \(\delta x, \delta y, \delta z\) from Gaussian distributions for each qubit with parameters described in Table \ref{tab:params} \cite{pagano2022error,wurtz2023aquila}. Decay parameters and all other relevant parameters for our training procedure and model are described in Table \ref{tab:params}.

\begin{figure}
    \begin{subfigure}[b]{0.23\textwidth}
        \centering
        \includegraphics[width=\textwidth]{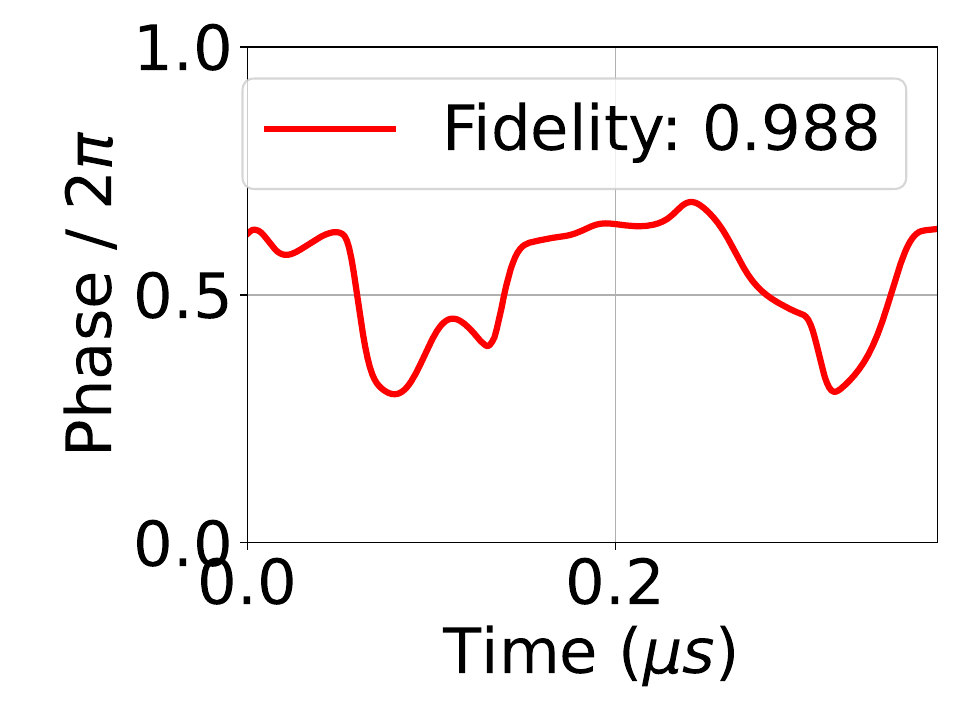}
        \caption{}
        \label{fig:pulse3_czz}
    \end{subfigure}'
    \hfill
    \begin{subfigure}[b]{0.23\textwidth}
        \centering
        \includegraphics[width=\linewidth]{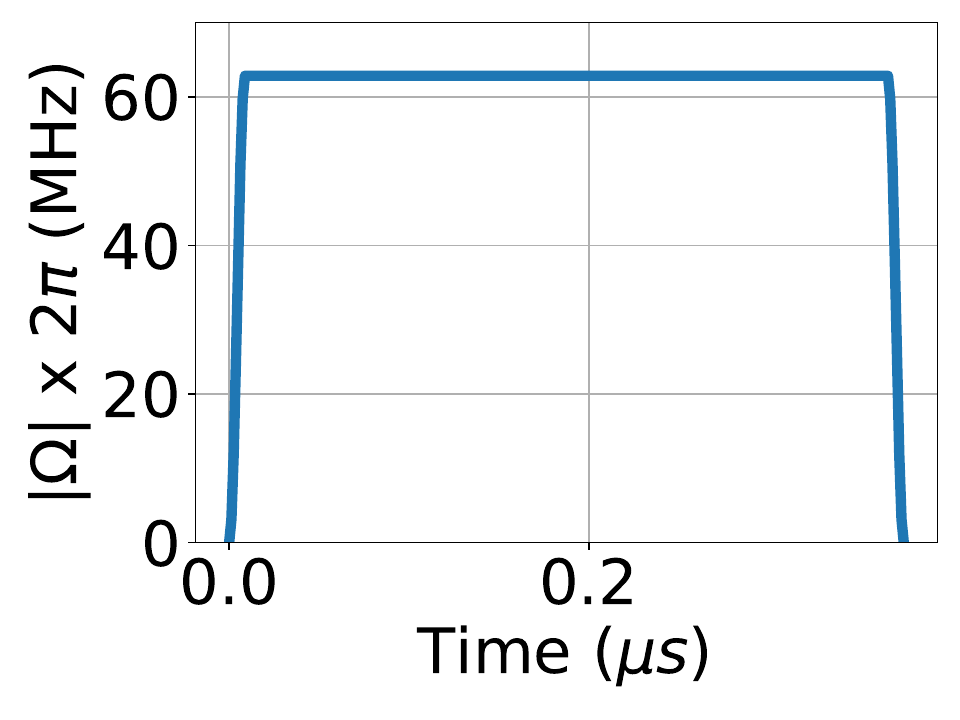}
        \caption{}
        \label{fig:pulse3_czz}
    \end{subfigure}
    \caption{(a) Phase pulse for $\mathrm{C}(Z^{\otimes 2})$ gate at $98.8\%$ fidelity (b) Omega drive with $10\mathrm{ns}$ ramp up, drive at $\Omega_\mathrm{Max}$, and ramp down for $10$ns}
    \label{fig:omega_drive}
\end{figure}

Combining the smoothness cost with an average infidelity across all sampled perturbations yields the final training loss:
\[
\mathcal{L}_\mathrm{total}
\;=\;
\underbrace{
   \bigl[\,
     1 - \langle \mathcal{F}\rangle_{\mathrm{batch}}
   \bigr]
}_{\substack{\text{average infidelity}\\\text{over displacements/decay}}}
\;+\;
\text{C}_{\text{smooth}},
\]
where \(\langle \mathcal{F}\rangle_{\mathrm{batch}}\) is the mean fidelity on a random set of perturbed Hamiltonians. We set our finite Rydberg lifetimes to $88\mu s$ \cite{evered2023high}. We set a gaussian ramp-up time to $\Omega_{Max}$ of $10ns$, describing the time to ramp up the Rabi drive to $\Omega_{Max}$ at the beginning of the pulse, and ramp down at the end of the gate \cite{evered2023high}. This is visualized in Figure \ref{fig:omega_drive}, and all $|\Omega|$ drives follow this pattern of ramp up - drive - ramp down.

\subsection{Pulse Results}

Based on the parameters outlined above, we discover realistic, high fidelity pulses that can implement the desired gate set. 

We define per-$\mathrm{C}Z$ error rate as the infidelity of a pulse, divided by the number of targets, specifically Error\_Per\_$\mathrm{C}Z$=$ \frac{1-\mathrm{Fidelity}}{N}$.

We demonstrate up to $99.55\%$ fidelity $\mathrm{C}(Z^{\otimes 2})$ gates, as seen in \ref{fig:czz_results_real}, resulting in a per-$\mathrm{C}Z$ error rate of $0.225\%$. The pulse profiles discovered are smooth, and performant under realistic atom motion. 

As for $\mathrm{C}(Z^{\otimes 3})$ gates, we demonstrate up to a $99.24\%$ fidelity pulse in \ref{fig:czzz_results_real}, resulting in a per-$\mathrm{C}Z$ error rate of $0.253\%$. We observe that the pulses learned for $\mathrm{C}(Z^{\otimes 3})$ gates are more complex and challenging to implement in real hardware, with the resultant pulses having sharper changes in phase when compared to the $\mathrm{C}(Z^{\otimes 2})$ gate. 

Implementing these optimized pulse shapes, particularly the phase modulations shown in Fig.~\ref{fig:czz_results_real} and Fig.~\ref{fig:czzz_results_real}, requires precise control over the driving laser field. While the smoothness constraint ensures the absence of instantaneous jumps, faithfully implementing over the gate duration ($\sim$250-600~ns) necessitates arbitrary waveform generators (AWGs) with sufficient bandwidth and resolution. Evered et. al. \cite{evered2023high} make use of a Spectrum M4i.6631-x8 AWG, which demonstrates that hardware specifications sufficient for implementing our simulated pulses are utilized in state-of-the-art experiments.

We report a set of $\mathrm{C}(Z^{\otimes 2})$ and $\mathrm{C}(Z^{\otimes 3})$ gate fidelities, times, and interatomic radii in Table \ref{tab:refined_fidelity} based on parameters in \ref{tab:params}. Corresponding pulses for various pulses are visualized in \ref{fig:czz_results_real} and \ref{fig:czzz_results_real}, with the Rabi drive amplitude from \ref{fig:omega_drive}-(b).

These pulses are evaluated with Rydberg decay, atom motion, and target real control constraints. The resultant pulses discovered have time scales of hundreds of nanoseconds, comparable with state of the art gate durations \cite{evered2023high}. These pulses allow for not only single effective $\mathrm{C}Z$ gate fidelities of as low as $0.20\%$, but also can enable reductions in circuit depth by expanding the underlying gateset. These pulse fidelities compete with current state of the art fidelities per-$\mathrm{C}Z$ of $\approx 0.5\% $ \cite{evered2023high}. These pulses align well with syndrome extraction, and can potentially be used to reduce syndrome extraction depth over various error correcting codes.

\subsection{Experimental Considerations}
\label{sec:exp_considerations} % Optional label for cross-referencing

While our simulations demonstrate the theoretical potential for high-fidelity C($Z^{\otimes N}$) gates (N=2, 3) using geometric blockade engineering under modeled noise, realizing these gates experimentally requires careful consideration of several practical challenges.

Our scheme relies fundamentally on precise relative positioning between the central control atom and the perimeter target atoms. The simulations incorporated static position errors sampled from Gaussian distributions with standard deviations $\sigma_{x} \approx \sigma_y \approx 0.014\,\mu$m and $\sigma_z \approx 0.16\,\mu$m (Table~\ref{tab:params}, based on \cite{pagano2022error,chew2022ultrafast}. Achieving and, crucially, maintaining such precision across multiple atoms within an array over the course of long computations is demanding, requiring highly stable trapping potentials and robust calibration routines to compensate for drifts, thermal effects, or differential light shifts \cite{scholl2021quantum, ebadi2021quantum, browaeys2020many}. Additionally, ensuring sufficient uniformity of the global laser pulse's intensity and phase across all participating atoms within the entangling region is important, as variations can introduce position-dependent errors \cite{bluvstein2024logical}.

% By surpassing the fidelity threshold often demanded in fault‐tolerant protocols, such as $0.8\%$ in Surface Code \cite{fowler2012surface}, these higher‐weight entangling gates open new opportunities for QEC transpilation, notably for multi-qubit stabilizer checks. They also avoid excessive atom rearrangements, a major time cost on neutral-atom devices. Combined, these features underscore that asymmetric blockade geometries—augmented with robust pulse design—can provide an avenue for scalable, multi-target controlled operations on single-species platforms, potentially transforming how QEC circuits are implemented at scale.

\section{Conclusion}

We have shown that multi-target, single-control gates $\mathrm{C}(Z^{\otimes 2})$ and $\mathrm{C}(Z^{\otimes 3})$ can be achieved on a single-species neutral-atom platform in one pulse by deliberately breaking the usual spatial symmetry of the Rydberg blockade. Positioning a control atom at the center and spreading target atoms on a sufficiently large perimeter balances strong control–target coupling against minimal target–target interactions. Our optimized global pulses are robust against Rydberg decay and random atom displacements, achieving gate fidelities up to $\sim 99.55\%$ for $\mathrm{C}(Z^{\otimes 2})$ and $\sim99.24\%$ for $\mathrm{C}(Z^{\otimes 3})$. While the specific circular geometry explored here faces inherent challenges in maintaining the required blockade asymmetry for N > 3 targets using a single pulse, the demonstrated $\mathrm{C}(Z^{\otimes 2})$ and $\mathrm{C}(Z^{\otimes 3})$ gates already represent valuable primitives for quantum computing, particularly for QEC.

In the future, it would be interesting whether analytical solutions to our pulse sequences could be found and whether our scheme could be generalized to larger $N$. In addition, other platforms with Rydberg interactions could make use of our framework, for example donors in silicon~\cite{Crane2019,craneRydbergEntanglingGates2021}, trapped ions~\cite{zhang2020}, or even fermionic neutral atom quantum computing platforms~\cite{gonzalez-cuadra2023,schuckert2024}. Such pulse optimisation could also be used for simulating fermionic systems, where multi-qubit gates appear in the form of Jordan-Wigner strings and which have been simulated in trapped-ion platforms~\cite{hemery2024} which have been shown to be able to implement multi-qubit gates~\cite{katz2023}.

\section*{Acknowledgements}

This research was supported by PNNL’s Quantum Algorithms and Architecture for Domain Science (QuAADS) Laboratory Directed Research and Development (LDRD) Initiative. The Pacific Northwest National Laboratory is operated by Battelle for the U.S. Department of Energy under Contract DE-AC05-76RL01830. This research used resources of the Oak Ridge Leadership Computing Facility, which is a DOE Office of Science User Facility supported under Contract DE-AC05-00OR22725.  A.S. acknowledges support from the U.S. Department of Energy, Office of Science, National Quantum Information Science Research Centers, Quantum Systems. This research used resources of the National Energy Research Scientific Computing Center (NERSC), a U.S. Department of Energy Office of Science User Facility located at Lawrence Berkeley National Laboratory, operated under Contract No. DE-AC02-05CH11231.

\bibliographystyle{quantum}
\bibliography{refs}

\newpage
\appendix 
\section{Pauli Transfer Maps\label{app:A}} % You can use \section or \subsection

\begin{figure*}[htbp] % htbp = here, top, bottom, page - placement suggestion
\centering
% *** REPLACE 'ptm_figure_1.png' with the actual filename of your first PTM plot ***
\includegraphics[width=2\columnwidth]{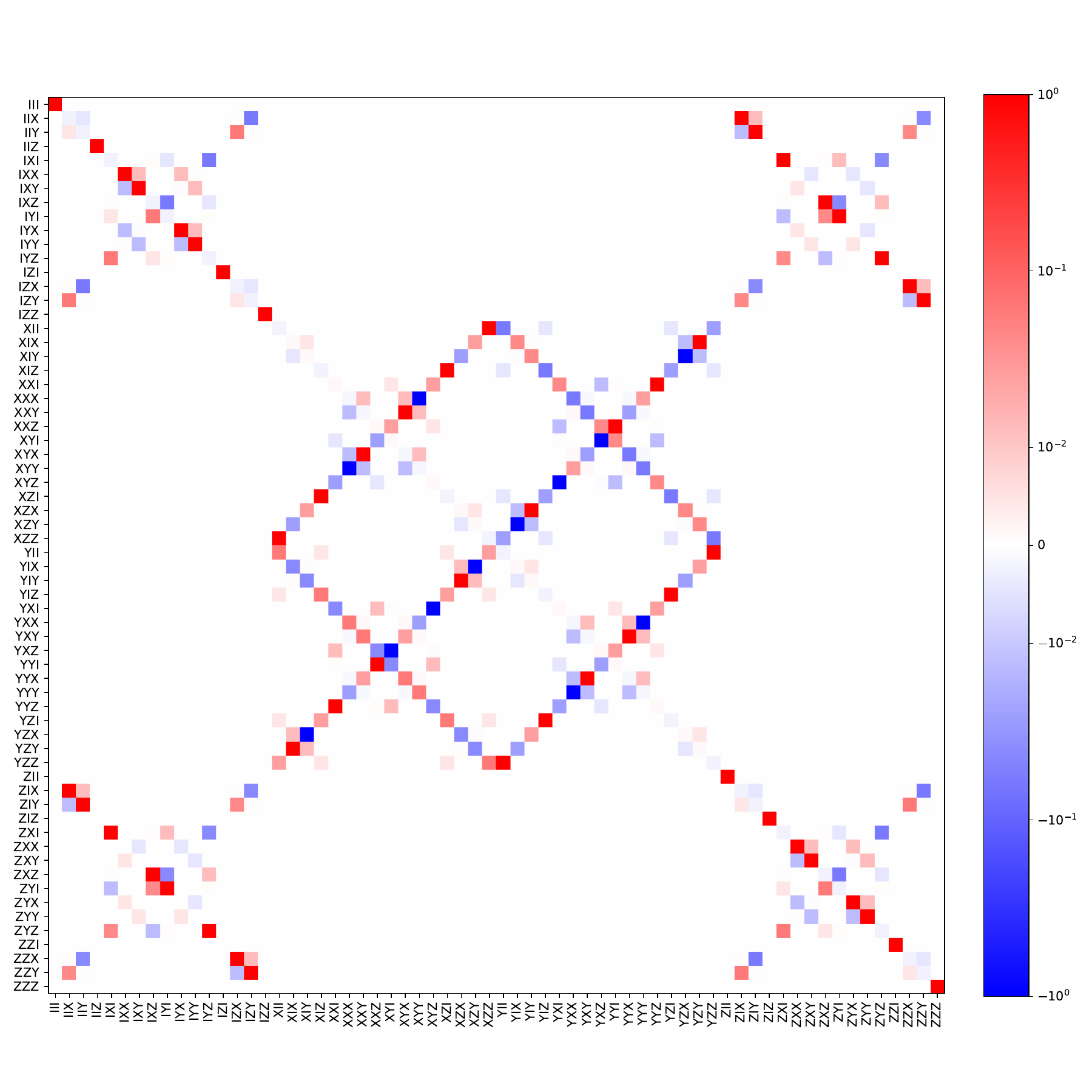} 
\caption{Logarithmic plot of the average Pauli Transfer Map (PTM) from simulation for the learned $\mathrm{C}(Z^{\otimes 2})$ gate. This map visualizes how input Pauli operators (indexed on the Y-axis) transform into output Pauli operators (indexed on the X-axis) under the learned quantum channel $\mathcal{E}$. The magnitude (color/intensity) of the PTM element $\mathcal{R}_{ij}$ indicates the strength of the $P_i \rightarrow P_j$ mapping. Lightly-shaded regions highlight the predominant error channels deviating from the ideal $\mathrm{C}(Z^{\otimes 2})$ operation, primarily identified as $X\leftrightarrow Y$ crosstalk and $Z\leftrightarrow I$ residual Rydberg population.}
\label{fig:ptm_2} % Use this label to refer to the figure in the text
\end{figure*}

\begin{figure*}[htbp] % htbp = here, top, bottom, page - placement suggestion
\centering
% *** REPLACE 'ptm_figure_1.png' with the actual filename of your first PTM plot ***
\includegraphics[width=2\columnwidth]{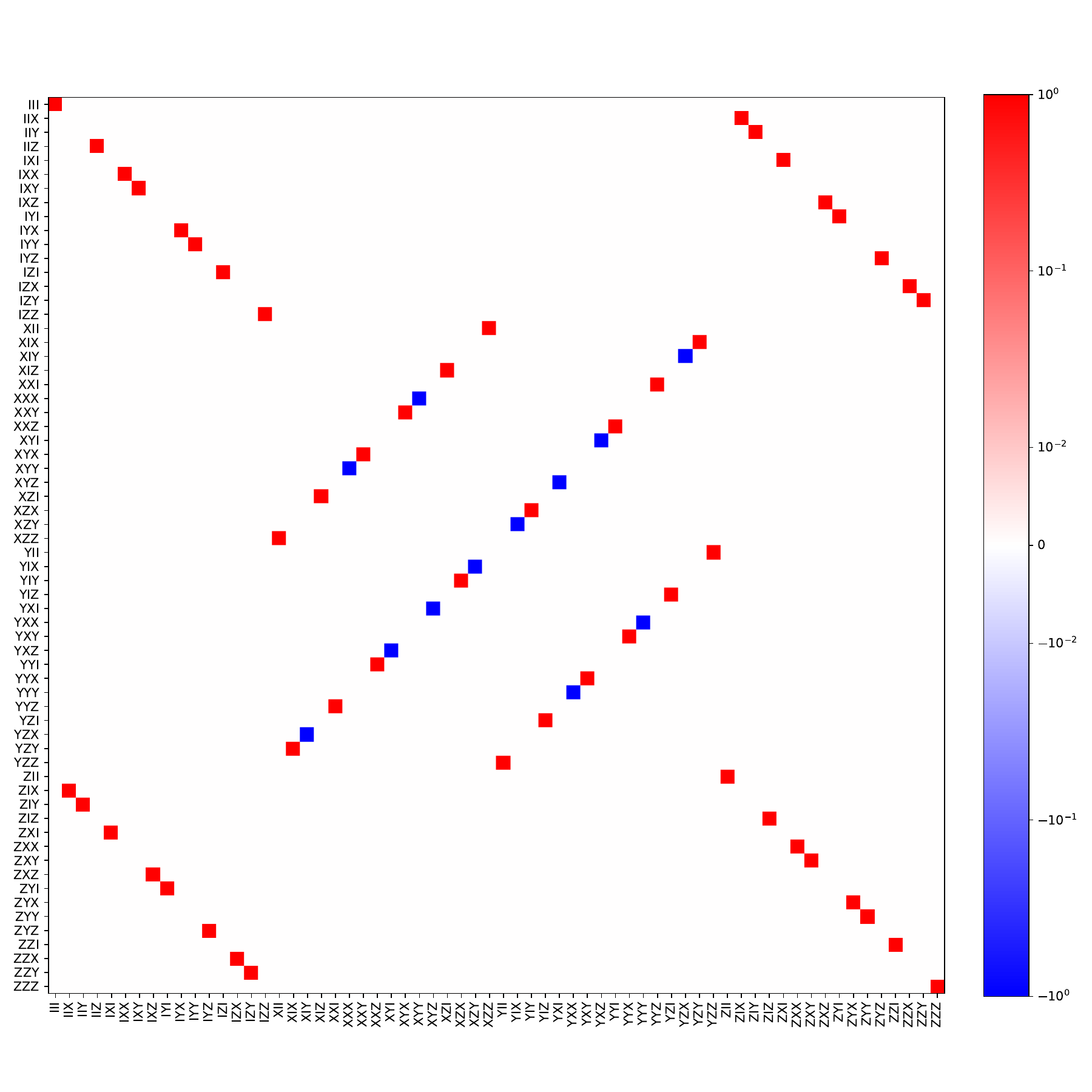} 
\caption{Theoretical Pauli Transfer Map (PTM) for an ideal, error-free $\mathrm{C}(Z^{\otimes 2})$ gate. This map illustrates the perfect transformation where input Pauli operators (Y-axis) map deterministically to output Pauli operators (X-axis). The PTM is sparse, containing only +1 or -1 entries at locations defined by the unitary $\mathrm{C}(Z^{\otimes 2})$ evolution, serving as a noiseless reference.}
\label{fig:ptm_1} % Use this label to refer to the figure in the text
\end{figure*}

Here we provide the Pauli Transfer Maps (PTMs) discussed in the main text. We compare the PTM learned without motion or decay (Fig.~\ref{fig:ptm_2}) and the ideal one (Fig.~\ref{fig:ptm_1}).

% Figure 1 in Appendix

% Figure 1 in Appendix

\end{document}